\documentstyle[psfig,12pt,a4wide]{article}

\newcommand{\news}{\setcounter{equation}{0}}

\def\eqn{\begin{equation}}
\def\eeqn{\end{equation}}
\def\arr{\begin{array}}
\def\earr{\end{array}}
\def\eqna{\begin{eqnarray}}
\def\eeqna{\end{eqnarray}}
\def\a{\alpha}
\def\b{\beta}

\def\D{\Delta}
\def\s{\sigma}

\def\w{\wedge}

\def\k{\kappa}
\def\O{\Omega}

\def\e{\epsilon}

\def\la{\lambda}
\def\La{\Lambda}

\def\t{\tau}
\def\p{\partial}

\font\mybb=msbm10 at 12pt
\def\bb#1{\hbox{\mybb#1}}
\def\bZ {\bb{Z}}
\def\bR {\bb{R}}
\def\bC {\bb{C}}
\def\bE {\bb{E}}

\def\bM {\bb{M}}

\def\bL {\bb{L}}

\def\spa#1{\phantom{\fbox{\rule[-#1cm]{0cm}{0cm}}}}

\begin{document}

\vspace*{-.6in} \thispagestyle{empty}
\begin{flushright}
ITFA--2002--07\\
LPTENS--02--19\\
\end{flushright}
\vspace{.2in} {\Large
\begin{center}
{\bf A New Cosmological Scenario in String Theory}
\end{center}}
\vspace{.2in}
\begin{center}
Lorenzo Cornalba$^{\dagger}$\footnote{lcornalb@yukawa.wins.uva.nl}\ \ 
and\ \ Miguel S. Costa$^{\ddagger}$\footnote{miguel@lpt.ens.fr}
\footnote{On leave from Departamento de F\'\i sica, 
Faculdade de Ci\^encias, Universidade do Porto.}
\\
\vspace{.2in} $^{\dagger}$\emph{Instituut voor Theoretische Fysica,
Universiteit van Amsterdam\\Valckenierstraat 65, 1018 XE Amsterdam, The Netherlands}\\
\vspace{.1in}
$^{\ddagger}$\emph{Laboratoire de Physique Th\'eorique de
l'\'Ecole Normale Sup\'erieure\\
24 rue Lhomond, F--75231 Paris Cedex 05, France}
\end{center}

\vspace{.3in}

\begin{abstract}
We consider new cosmological solutions with a collapsing, an
intermediate and an expanding phase. The boundary between the
expanding (collapsing) phase and the intermediate phase is seen by
comoving observers as a cosmological past (future)
horizon. The solutions are naturally embedded in string and M--theory. 
In the particular case of a
two--dimensional cosmology, space--time is flat with an identification 
under boost and translation transformations. We consider
the corresponding string theory orbifold and calculate the 
modular invariant one--loop partition function. In this case there is
a strong parallel with the BTZ black hole. The higher dimensional
cosmologies have a time--like curvature singularity in the
intermediate region. In some cases the string coupling can be made
small throughout all of space--time but string corrections become
important at the singularity. This happens where string
winding modes become light which could resolve the singularity. 
The new proposed space--time causal structure could have implications 
for cosmology, independently of string theory.
\end{abstract}
\newpage

\section{Introduction}

One of the central problems in our present view of the Universe has to
do with the cosmological singularity. The observation of an
expanding Universe leads us to believe that in the past the Universe was
much denser. At the Planck scale General Relativity breaks down where it is usually
believed that a space--like cosmological singularity develops. Despite the
great advances of particle cosmology from the GUT scale to present times, the
understanding of the cosmological singularity remains a challenge. 
The resolution of this problem is one of the main motivations to find a 
quantum theory of gravity.

It as long been understood that string theory, as a consistent theory
of gravity, could be a good starting point to investigate the universe
at the Planck scale (see \cite{Veneziano:00} for a review and
references). In the pre Big--Bang scenario, the Universe starts in a
contracting phase until string effects along a space--like hypersurface 
become important. From this space--like hypersurface the Universe will
evolve to the present expanding phase. The main problem that has prevented a deeper 
understanding of the singularity problem is the understanding of the stringy phase. 
Recently, there has been a new proposal for a Big--Crunch/Big--Bang transition
\cite{Khoury:01,Seiberg:02}. These authors considered a toy model in
which space--time is seen  as flat space
with an identification along boost transformations, an orbifold earlier
investigated in \cite{HoroSteif:91}. A problem in this string
compactification is that there are closed time--like curves and
space--time is not Hausdorff. So even if space--time is flat and there
is no curvature singularity, there is still a singularity at the
Big--Bang. This problem has motivated our research. We shall present
a two--dimensional toy model where we manage to smooth this string
orbifold and to hide the closed time--like curves behind a
cosmological horizon. Related work, with a different kind of orbifold,
appeared recently in \cite{Balasubramanian:02}.

The two--dimensional toy model may have a tractable string theory
description as an orbifold of flat space, and we initiate its
investigation. On the other hand, as a solution of the gravity low
energy equations, one expects that the new space--time causal structure of the
two--dimensional toy model can be generalized to higher
dimensions. Indeed, we shall construct new cosmological solutions for
open Universes in arbitrary dimensions with three
phases: (1) A collapsing phase with a future cosmological horizon; (2)
An intermediate phase where there is a time--like curvature 
singularity; (3) An expanding phase with a past cosmological horizon. 
In the context of string theory, we shall argue that near the time--like
curvature singularity there are string winding modes that are becoming
very light. This fact, led us to conjecture that when these new light
states are taken into account the singularity could be resolved,
smoothing the geometry. 

This paper is organized in the following way. In section 2 we present
the new space--time identifications that lead to a two--dimensional
cosmological solution. Here we were inspired by an analogous
construction for the BTZ black hole \cite{Banados:92,Banados:93}. We start with
three--dimensional flat space and identify points in
space--time under boost and translation transformations. After 
Ka\l u\.{z}a--Klein reduction, we find a cosmological solution with the
properties described in the previous paragraph. In this case,
however, there is no singularity. The apparent singularity
corresponds to the surface where the compactification circle becomes
null. As a consequence, there will be causal time--like curves in
the intermediate region that can be closed if one passes the
causal singularity. The cosmological observers in the expanding and
collapsing regions will not intersect these closed time--like curves.

The geometrical construction of the two--dimensional
model can be embeded in string theory. This topic is
covered in section 3. We consider the case of a  string
orbifold and analyse first the region of validity of the corresponding
gravity solution. It is seen that in the intermediate region, where there
are closed time--like curves, string effects become important. We
calculate the one--loop bosonic partition function for the orbifold and show
that it is modular invariant. The introduction of a
translation in the boost identification, not only makes space--time
smoother, but also introduces a regularization scheme in the
calculation of the string partition function. 

In section 4 we generalize the two--dimensional cosmological toy model
to higher dimensions. We start with a theory of gravity with a scalar
field and a form field strength. Then we compactify
space--time on a (Ricci) flat manifold along which
there is a flux of the form field. Comparison with the
two--dimensional case led us to require the geometry to be smooth and
invariant along a null hypersurface that we interpret as the
cosmological horizon. These boundary conditions and the homogeneity
requirement of the cosmological solution, imply that the geometry has
the form of the higher dimensional Milne Universe along this null
hypersurface, leading to a negative curvature Universe. 
We analyse the resulting geometry, which has
collapsing, intermediate and expanding regions as mentioned above. 
We finish this section by embedding the new solutions in string and
M--theory. Related work that considers a different type of
time--dependent string theory solutions by fixing boundary conditions
on space--like hypersurfaces appeared very recently in \cite{GutpStro:02}. 

In section 5 we give our conclusions. We show that in the new
cosmological solutions the usual cosmology horizon problem that not
arise. Then we comment on the possible resolution of the
cosmological time--like singularity. We argue that many new solutions
can be found by imposing the boundary conditions using a different
scalar potential. In particular, one can consider potentials of the type
used in standard particle cosmology, avoiding the space--like
singularity of the Big--Bang. 

\section{Cosmological solution as a quotient of flat space}
\news

Consider flat space--time in $2+1$ gravity with line element
\eqn
ds^2=-dT^2+dX^2+dY^2\ .
\label{flatmetric}
\eeqn
We shall identify points on this space along orbits of a subgroup of
its isometry group, i.e. a subgroup of the three--dimensional
Poincar\'e group. The situation is analogous to the BTZ black hole 
that is obtained from $AdS_3$ in $2+1$ gravity with a negative
cosmological constant \cite{Banados:92,Banados:93}. Later we shall embed
the three--dimensional geometry here derived in string and M-theory. 

Let $\k$ be the Killing vector 
\eqn
\k =2\pi i \left(\D J + R\,P\right)\ ,
\label{kappa}
\eeqn
where 
\eqn
iJ=T\frac{\p}{\p X}+X\frac{\p}{\p T}\ ,\ \ \ \ \ \ \ \ 
iP=\frac{\p}{\p Y}\ ,
\eeqn
are the generators of Lorentz boosts along the $X$ direction and
translations along the $Y$ direction, respectively. The Killing vector
$\k$ defines a one parameter subgroup of isometries. We shall identify
points $Q$ along the orbits of this subgroup according to
\eqn
Q \sim \exp{(\k)}\, Q\ .
\eeqn
The parameter $\D$ is related to the boost velocity by
$v=\tanh{(2\pi\D)}$ and $2\pi R$ is the translation length. The
resulting space is a smooth manifold with a flat metric, because
the identifications are along an isometry of the initial space with no
fixed points. However, now there are curves that join identified points and
one needs to worry about the causal structure  of the resulting space. 

A necessary condition for the absence of time--like and null curves is
$\k\cdot\k > 0$. The boundary of this region is the surface
$\k\cdot\k=0$ described by
\eqn
-T^2+X^2=
\frac{1}{E^2}\ ,
\label{kappanull}
\eeqn
where $E$ has dimension length$^{-1}$ and is defined by
\eqn
E=\frac{\D}{R}\ .
\label{E}
\eeqn
Notice that the action induced by the Killing vector $\k$ has no fixed
points because the vector $P$ always induces a translation along the
$Y$ direction. The orbits of $\k$ in the quotient space are closed
curves and cannot be continuously deformed to a point. These orbits are
null on the surface defined by (\ref{kappanull}) and become time--like
beyond it. 

Recently, Khoury et al \cite{Khoury:01} considered the
cosmological solution arising from the identification of points in
space--time that are related by a boost transformation. This case
corresponds to setting $R=0$ in our model. In their work, the point
$T=X=0$ is a fixed point of the orbifold and the light rays $|T|=|X|$
are mapped arbitrarily close to the origin. As a consequence, if one
does not excise any region of space-time, there will be causal closed
curves arbitrarily close to the point $T=X=0$. Also, 
the resulting manifold is not Hausdorff. 
In the model here proposed, the introduction of the
translation in the space--time identification has resolved these
problems. In particular, we also have causal closed curves but 
we shall argue that, from the point of view of the
cosmological observer, these curves are not observable, and no
inconsistencies should arise. 

To understand the causal structure of the quotient space it is
convenient to divide space in three different regions
\eqn
\arr{rcl}
I&:&|T|>|X|\ \ {\rm and}\ \ \k\cdot\k>0\ ,
\\
II&:&|T|<|X|\ \ {\rm and}\ \  \k\cdot\k>0\ ,
\\
III&:&\k\cdot\k<0\ .
\earr
\label{regions}
\eeqn
We shall call region $I$ the outer--region. For $T>0$ ($T<0$) it
describes an expanding (collapsing) universe. In this region, the
above condition can be resumed to $\k\cdot\k>(2\pi R)^2$. 
Region $II$ is the inside--region, where $0<\k\cdot\k<(2\pi R)^2$. 
The region beyond which $\k$ becomes time--like is called region $III$.
In figure 1 the different regions are
represented. The frontiers between regions $I$ and $II$ are
null surfaces. This fact is very important because these surfaces
become horizons and prevent causal closed curves to be extended to
both the expanding and collapsing outer regions. 

\begin{figure}
\begin{picture}(0,0)(0,0)
\end{picture}
\centering\psfig{figure=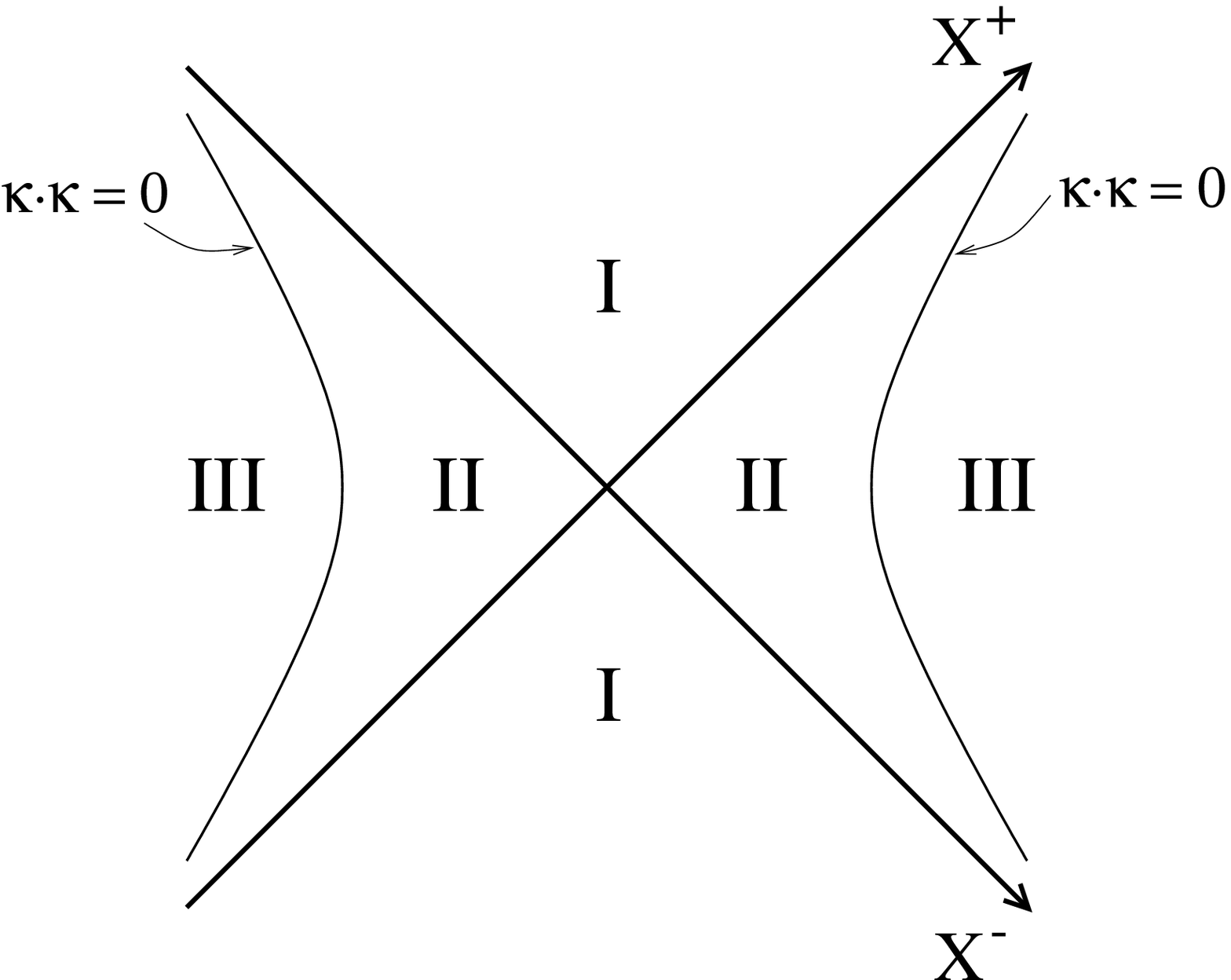,width=7cm} 
\caption{\small{The different regions in space--time in the light cone
coordinates $X^{\pm}=(X\pm T)/\sqrt{2}$. The outer
regions $I$ will be interpreted as the cosmological collapsing 
and expanding phases. In regions $III$ there are closed time--like
curves that can be deformed to regions $II$, but always close in
region  $III$. The surface $|T|=|X|$ acts as a horizon because the
closed time--like curves cannot be deformed to enter regions $I$.}}  
\label{fig1}
\end{figure}

\subsection{Compactification to two dimensions}

We want to interpret this flat geometry from the two--dimensional
point of view. In other words, we want to find the coordinate
transformation that brings the Killing vector $\k$ to the form 
\eqn
\kappa= 2\pi R\,\frac{\p}{\p y}\ .
\label{newkappa}
\eeqn
Then, to obtain the two--dimensional cosmological solution we consider 
the Ka\l u\.{z}a--Klein compactification
\eqn
ds_{3}^{\ 2}=ds_2^{\ 2}+e^{2\s}\left( dy+A_adx^a\right)^2\ ,
\eeqn
keeping in mind that one can always add extra spectator dimensions.
The above Ka\l u\.{z}a--Klein reduction of the three--dimensional
Einstein--Hilbert action gives
\eqn
S=\frac{1}{2\kappa_2^{\ 2}}\int d^2x\sqrt{-g}e^{\s}
\left[R -\frac{1}{4}e^{2\s} F^2\right]\ ,
\eeqn
where $\kappa_2$ is the two--dimensional gravitational coupling.
The only dynamical degree of freedom is the scalar field.

Now let us analyse the geometry from the
two--dimensional point of view. We consider the following coordinate
transformations in both regions $I$ and $II$
\eqn
I:\left\{
\arr{l}
T=t\cosh{[E(x+y)]}\\
X=t\sinh{[E(x+y)]}\\
Y=y
\earr
\right.\ ,\ \ \ \ \ \ \ \ \ 
II:\left\{
\arr{l}
T=t\sinh{[E(x+y)]}\\
X=t\cosh{[E(x+y)]}\\
Y=y
\earr
\right.\ .
\label{coordtransf}
\eeqn
In the new coordinate system the killing vector $\k$ as the form
(\ref{newkappa}) required for compactification. We consider first the
region $I$, where in the new coordinate system the line element becomes
\eqn
ds^2=-dt^2+\frac{(Et)^2}{\La(t)}dx^2+
\La(t)\left(dy+\frac{(Et)^2}{\La(t)}dx\right)^2\ ,
\label{cosmosol}
\eeqn
with
\eqn
\La(t)=1+(Et)^2\ .
\label{lambda}
\eeqn
From the above Ka\l u\.{z}a--Klein ansatz it is straightforward to read
the two--dimensional fields from the metric element. Both the $t$ and
$x$ coordinates run from $-\infty$ to $+\infty$. The region $t<0$
describes a contracting universe while the region $t>0$ describes an
expanding universe. The time coordinate $t$ is the proper time for a
comoving observer with the expansion (or collapse).  We shall focus on
the expanding region. The Ka\l u\.{z}a--Klein 2--form field strength
is given by 
\eqn
F=\frac{2E^2t}{\La^2}\, dt\w dx\ ,
\label{electfield}
\eeqn
which vanishes for large cosmological times, where the metric becomes
flat. The radius of the compactification circle $R(t)=R\sqrt{\La(t)}$,
determined by the scalar field, is growing with time, and therefore, at
later times the Ka\l u\.{z}a--Klein approximation is no longer
valid. At earlier times, for $Et\ll 1$, the metric becomes 
\eqn
ds^2\sim -dt^2 + (Et)^2dx^2 \ .
\eeqn
At the surface $t=0$, $x=\pm\infty$ there is a harmless coordinate
singularity. In fact, at this surface the metric has the usual form of the Milne
universe obtained from a coordinate transformation of the usual flat
space metric. This surface represents the past cosmological horizon
for the observers in the expanding region $I$. The radius of the compact
circle on the horizon is $R$. 

The Ka\l u\.{z}a--Klein form of the metric in region $II$ can be obtained
from the coordinate transformation (\ref{coordtransf}). The final result
is a line element that can be obtained from that of region $I$ in
(\ref{cosmosol}) by the replacement $t\rightarrow it$
\eqn
ds^2=-\frac{(Et)^2}{\La(t)}dx^2+dt^2+
\La(t)\left(dy-\frac{(Et)^2}{\La(t)}dx\right)^2\ ,
\label{cosmosolII}
\eeqn
where
\eqn
\La(t)=1-(Et)^2\ .
\label{lambdaII}
\eeqn
Now the coordinates $t$ and $x$ have fliped their spatial and time--like
caracter. The surface $t=0$, $x=\pm\infty$ becomes a horizon for
an observer in region $II$ using this coordinate system. In fact, near
this surface the metric becomes the Rindler metric.
As we move away from the horizon the compactification radius 
decreases and vanishes at $Et=1$. At this surface the two--dimensional
curvature and the invariant $F^2$
diverge. This is the surface where the killing vector $\k$ becomes
null, and the Ka\l u\.{z}a--Klein approximation breaks down. 
Of course from the higher dimensional point of view this is not a
curvature singularity because the metric is a quotient of flat space,
but it is the locus where the orbits of $\k$ become causal. Figure 2
contains the Carter--Penrose diagram for the geometry at a fixed position in
the compact direction $y$. Similar two--dimensional space--time causal
structures were considered in \cite{Kounnas:92,Grojean:01}.

\begin{figure}
\begin{picture}(0,0)(0,0)
\put(145,230){${\cal I}^+$}
\put(70,30){${\cal I}^-$}
\end{picture}
\centering\psfig{figure=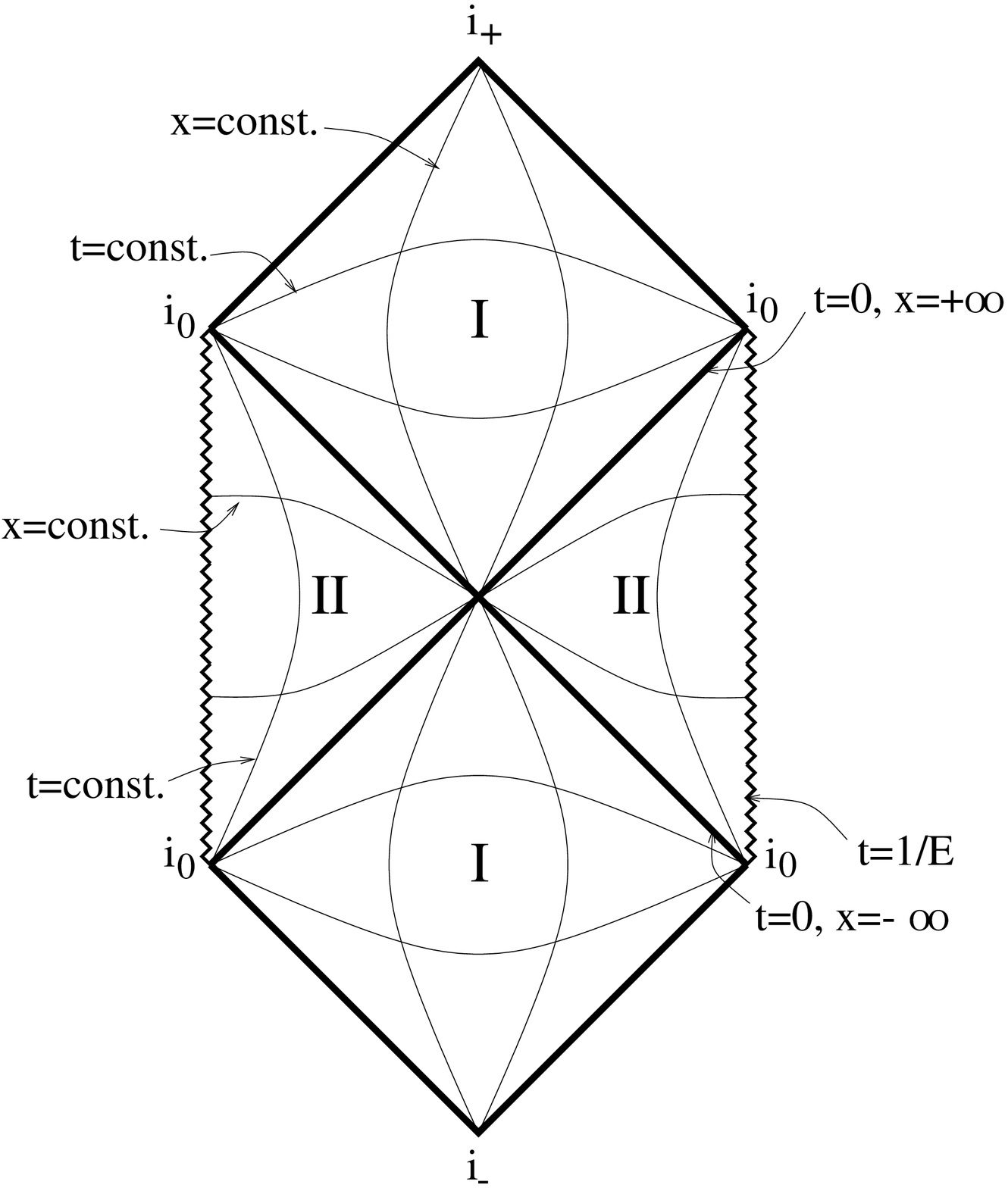,width=8cm} 
\caption{\small{Carter--Penrose diagram for the two--dimensional 
Ka\l u\.{z}a--Klein cosmology. Spatial, future and past infinities are 
defined with respect to the expanding region $I$. The past
(future) horizon of the expanding (collapsing) outer region is a
Cauchy surface. The singularity at $t=1/E$ is an artifact of the
compactification because this is the surface where $\k$ becomes
null.}}  
\label{fig2}
\end{figure}

As a remark let us note that there is a set of coordinates that
extends the metric in the outer-region $I$ to the inner--region
$II$, and that this metric is well behaved throughout these
regions. This is easily seen because the metric in region $I$ near the
horizon behaves exactly as that for the Milne universe. So the
coordinate transformation  
\eqn
\tilde{T}=t\cosh{(Ex)}\ ,\ \ \ \ \ \ \ 
\tilde{X}=t\sinh{(Ex)}\ ,
\label{tildeTX}
\eeqn
in the solution (\ref{cosmosol}) also covers region $II$. 

\subsection{Geometry as a limit of the BTZ black hole}

We have described, in the previous subsections, geometries derived as
quotients of flat Minkowski three--dimensional space. In this
subsection we show that these geometries have natural generalizations
as quotients of $AdS_{3}$, which are nothing but the usual BTZ black
hole solutions. Therefore we can regard the geometries considered
previously as a limit of the BTZ black hole.

To make this relation more precise, we describe $AdS_{3}$, as usual, as
the surface 
\eqn
-\left( Z^{0}\right)^{2}+\left( Z^{1}\right)^{2}
+\left( Z^{2}\right)^{2}-\left( Z^{3}\right) ^{2}=-L^{2}\ ,
\eeqn
with the parameterization 
\eqn
\arr{rcl}
Z^{0}+iZ^{3} &=&-i\,e^{\frac{iT}{L}}\sqrt{L^{2}+X^{2}+Y^{2}}\ , \\
Z^{1}+iZ^{2} &=&X+iY\ .
\earr
\eeqn
The symmetry group of $AdS_{3}$ is $SO\left( 2,2\right) $, with generators 
$iJ_{\mu \nu }=Z_{\mu }\partial _{\nu }-Z_{\nu }\partial _{\mu }$. In the
limit $L\rightarrow \infty $ we recover flat Minkowski space
$\bR^{1,2}$ with isometry group $ISO\left( 1,2\right)$. 
Moreover, the generators $J_{\mu \nu }$ converge to the generators 
$P_{m}$, $M_{mn}$ ($m,n=0,1,2$) of $ISO\left( 1,2\right)$ according to  
\eqn
J_{mn}\rightarrow M_{mn}\ ,\ \ \ \ \ \ \ \ \ \ \ \ \ 
\frac{1}{L}\,J_{3m}\rightarrow P_{m}\ .
\eeqn
Therefore, the natural generalization of the construction considered in the
last section is to consider the quotient of $AdS_{3}$ by the isometry 
$e^{\kappa}$, where 
\eqn
\kappa =2\pi i
\left[\Delta J_{10}+\left( \frac{R}{L}\right) J_{32}\right]\ .
\eeqn
This corresponds \cite{Banados:93} to the BTZ black hole geometry \cite{Banados:92}
\eqn
ds_{BTZ}^{2}=-N^{2}dv^{2}+N^{-2}dr^{2}+r^{2}\left(d\phi+N^{\phi}dv\right)^{2}\ ,
\eeqn
where 
\eqn
N^{2}=\frac{1}{L^{2}r^{2}}
\left(r^{2}-r_{+}^{2}\right)\left(r^{2}-r_{-}^{2}\right)\ ,
\ \ \ \ \ \ \ \ \ \ \ \ 
N^{\phi}=\frac{1}{L}\frac{r_{+}r_{-}}{r^{2}}\ ,
\eeqn
and 
\eqn
r_{+}=L\Delta\ ,\ \ \ \ \ \ \ \ \ \ \ r_{-}=R\ .
\eeqn
We assume that $L>E^{-1}$, so that $r_{+}>r_{-}$. 
Recall that this geometry describes a rotating black hole with inner and
outer horizons at $r=r_{-}$ and $r=r_{+}$. In the regions $r>r_{+}$ and $%
r<r_{-}$ the coordinate $r$ is space--like, while in the region $%
r_{-}<r<r_{+}$ the coordinate $r$ is time--like. 

The Euclidean continuation of the time coordinate, $v\rightarrow\t_E$,
in the regions $r>r_{+}$ and $r<r_{-}$, yields quotients of the
hyperbolic three--plane $H_{3}$ under an element of the symmetry group 
$SO\left( 1,3\right)$ \cite{Carlip:94}. Recall that quotients
of $H_{3}$ are parameterized by the upper complex half--plane as follows. If
we write the metric of $H_{3}$ as
\eqn
ds_{H_{3}}^{\ 2}=L^{2}\left[ \left( 1+\rho ^{2}\right) d\t_E^{\ 2}
+\frac{d\rho ^{2}}{1+\rho ^{2}}+\rho ^{2}d\phi^{2}\right]\ ,  
\label{eq1000}
\eeqn
then a quotient parameterized by $\tau=\tau_{1}+i\tau_{2}$ is given by the
identifications 
\eqn
\t_E\sim\t_E+2\pi \tau _{2}\ ,\ \ \ \ \ \ \ \ \ \ \
\phi\sim\phi+2\pi \tau _{1}\ .
\eeqn
This corresponds, from the point of view of the 2--dimensional CFT on the
$AdS$ boundary, to considering the CFT on a torus with modular parameter 
$\tau$ \cite{MaldaStro:98}. The Euclidean continuation of the
regions $r>r_{+}$ and $r<r_{-}$ give quotients of $H_{3}$ with
modular parameters $\tau_{\mathrm{out}}$ and $\tau_{\mathrm{in}}$,
respectively, where 
\eqn
\arr{c}
\displaystyle{\tau_{\mathrm{out}}=\frac{1}{L}\left( r_{-}+ir_{+}\right) 
=\frac{R}{L}+i\Delta}\ ,
\spa{0.4}\\
\displaystyle{\tau_{\mathrm{in}}=\frac{1}{L}\left( r_{+}+ir_{-}\right) 
=\Delta +i\frac{R}{L}}\ .\ 
\earr
\eeqn

Now we can easily connect the discussion above with the compactification
considered in the previous subsections, which is obtained in the limit 
$L\rightarrow \infty$, with $R,\Delta $ and $r$ fixed. In this limit
clearly $r_{-}=R$ remains finite and
\eqn
r_{+}=\Delta L\rightarrow \infty\ ,
\eeqn
so that the outer region $r>r_{+}$ is no longer part of the
geometry. Then the intermediate region $r>r_{-}$, corresponds to
region $I$ in the previous subsections, and $r$, which is a time--like
coordinate in this region, is related to the cosmological time. More
precisely, in the limit $L\rightarrow\infty$ one obtains the geometry
\eqn
ds_{BTZ}^{2}=-\frac{1}{\Delta^{2}}\left(\frac{r^{2}}{r^{2}-R^{2}}\right)
dr^{2}+\Delta^{2}\left(\frac{r^{2}-R^{2}}{r^{2}}\right)dv^{2}
+r^{2}\left(d\phi +\frac{\Delta R}{r^{2}}dv\right)^{2}\ ,
\eeqn
which is nothing but the solution (\ref{cosmosol}) under the change of
coordinates 
\eqn
\left(\frac{r}{R}\right)^2=1+\left( Et\right)^{2}\ ,
\ \ \ \ \ \ \ \ \
\Delta\,v=-x\ ,
\ \ \ \ \ \ \ \ \ 
R\,\phi=y+x\ .
\eeqn

In the above geometry one can consider the Euclidean continuation of the
inner region $r<r_{-}$, which, as described above, gives the quotient of $%
H_{3}$ described by the modular parameter $\tau _{\mathrm{in}}$. The
geometry (\ref{eq1000}) becomes (calling $L\t_E=z$ and $L\rho =r$) the usual
flat Euclidean metric $dz^{2}+dr^{2}+r^{2}d\phi^{2}$ with identifications
\eqn
z\sim z+2\pi R\ ,
\ \ \ \ \ \ \ \ \ \ 
\phi \sim \phi +2\pi \Delta\ . 
\eeqn
This is clearly the geometry of an Euclidean flux brane, as it is natural to
expect.

\subsection{Closed time--like curves?}

Now we come to the delicate point of wether we should excise
regions $III$, interpreting the surface $\k\cdot\k=0$ as a
causal singularity. We shall advocate that from the point of view of
an observer in the expanding (or collapsing) cosmological
region $I$, there is  no contradiction arising from the inclusion of
those regions. 

There are closed time--like curves in the region $III$ where  
$\k\cdot\k < 0$. These curves can be deformed to the region $II$,
resulting in causal closed curves that are partially in region $II$
(one needs the region $III$ to close the open causal curves
in region $II$ that start and end on the surface
$\k\cdot\k=0$). However, there are no causal closed curves passing
through region I because of the cosmological horizon. Indeed the
horizon is a Cauchy surface and therefore no causal curve intersects
it more than once. Hence the
comoving observer never intersects any 
closed causal curve. In this sense the existence of the closed
time--like curves is harmless. Any set of initial conditions at
cosmological time $t_0$ will evolve without contradiction. Also, the
existence of closed time--like curves shielded by horizons appears in
the BTZ black hole \cite{Banados:92,Banados:93}, as well as in other
higher dimensional rotating black holes \cite{Herdeiro:00}, both cases
having string theory dual descriptions. Furthermore, near the causal
singularity one expects new degrees of freedom to become important
which could prevent the existence of the closed time--like curves (see
\cite{Maldacena:01} for a related discussion in the case of the BTZ
black hole).
 
This geometry represents a smooth transition from a
collapsing phase to an expanding phase, with the additional regions
$II$ and  $III$ in between.
To understand how a set of initial boundary conditions propagates from
the collapsing to the expanding phase, it is necessary to understand
the evolution throughout the intermediate regions.
Therefore it is important to
investigate the modifications in field and string theories due to the
boundary conditions here imposed. For example, in region $II$ there is
an increasing electric field and one would expect that pair production of 
Ka\l u\.{z}a--Klein particles occurs. Also, the proper radius of the 
Ka\l u\.{z}a--Klein circle is decreasing and therefore these particles
become very massive as the electric field becomes very large.
We can do an estimation of the pair production rate using the Schwinger
formula $\Gamma\sim exp\left(-\pi\frac{m}{{\cal E}}\right)$, where
$m=1/R(t)$ is the particle's mass and 
${\cal E}^2=-F^2/2$ determines the electric field. 
This gives the following estimation for the pair production rate
\eqn
\Gamma\sim \exp{\left( -\frac{\pi\La(t)}{2ER}\right)}\ .
\eeqn
Asymptotically in region $I$ this gives a very small rate. Along the
light rays, it can be made small provided $ER=\D\ll 1$. In region 
$II$ the argument in the exponential becomes very small and the
semi--classical approximation for the nucleation rate breaks down. 
Nevertheless, one expects that there will be Ka\l u\.{z}a--Klein particles
produced in region $II$ that, 
for later cosmological times in region $I$,
move  at constant velocity and will be distributed homogeneously
through the constant time sections of the comoving observer with the
expansion. Note that this invariance under translations in $x$ is
nothing but the unbroken $SO(1,1)$ subgroup of the $ISO(1,2)$ left
invariant by the compactification.

Clearly, the above example points to a better understanding of
quantum processes in this space--time. In the next section, we
start the investigation of these issues by embedding this construction 
in string theory.

\section{String Orbifold as a Cosmology}
\news

The above construction of a cosmological background by a quotient of
flat space can be embedded in string and M--theory by adding the
appropriate number of flat spectator directions. In the case of
ten--dimensional string theory one has the usual string in flat space
but there is a twisted sector arising from the orbifold
compactification. Before we analyse this string theory let us briefly
describe the underlying geometry and its limits of validity. 

Considering type II strings, the compactification
on a circle to the nine--dimensional string frame is given by
\eqn
ds_{10}^{\ 2}=ds_9^{\ 2}+e^{2\s}\left( dx^9+A_adx^a\right)^2\ ,
\eeqn
where $\phi_9\equiv \phi -\s/2$ is the nine--dimensional dilaton
field. If one adds seven spectator flat directions to the construction
of the last section and reduces along the $y$ direction, one obtains
\eqn 
\arr{c} 
\displaystyle{ ds_9^{\ 2}=-dt^2+\frac{(Et)^2}{\La(t)}\,dx^2+ds^2(\bE^7)\ ,}
\\
\displaystyle{ e^{2\s}= 1+(Et)^2\equiv \La(t)\ ,\ \ \ \ \ 
A=\frac{(Et)^2}{\La(t)}\,dx\ .}  
\earr
\label{9Dsol}
\eeqn 
The space--time causal structure of this geometry and its maximal
extension was explained before. 
The conditions for the validity of the
nine--dimensional description are as follows. Firstly, the
nine--dimensional string coupling has to be  small 
\eqn
g_9=\frac{g\a'^{1/4}}{\sqrt{R(t)}}\ll 1\ .
\eeqn
In region $I$ the coupling decreases with time, therefore 
this condition always holds provided the ten--dimensional string 
coupling $g$ is small. In region $II$, the condition fails
near the causal singularity and one needs to use the ten--dimensional
approximation. Indeed, since the ten--dimensional
coupling $g$ remains constant, the string orbifold analysis holds as
long as $g$ is kept small.

Secondly, the typical energy scale 
$E=\D/R$ in the geometry described by (\ref{9Dsol}) should be
much smaller then the massive string states scale
\eqn
E\ll\frac{1}{\sqrt{\a'}}
\ \ \ \Rightarrow\ \ \ 
\D\ll\frac{R}{\sqrt{\a'}}\ .
\label{stringscale}
\eeqn
This condition can always be satisfied for $\D$ sufficiently
small. Thirdly, the string winding states should be very massive, i.e.
\eqn
E\ll\frac{R(t)}{\a'}\ ,
\eeqn
which in region $I$ is compatible with the condition above. However, in
region $II$ the winding states become very light because the proper
size of the circle converges to zero. This means that in this
region new light string degrees of freedom become important and one
needs to use the string orbifold description. Finally, space--time is
effectively nine--dimensional provided the Ka\l u\.{z}a--Klein modes
are very massive
\eqn
E\ll\frac{1}{R(t)}\ .
\eeqn
In region $I$, this condition will fail for large cosmological
time. However, the region of validity can be made arbitrarily large by
choosing $\D$ very small. In region $II$ the condition is satisfied
because the Ka\l u\.{z}a--Klein modes become very massive. At last,
one could worry that the curvature corrections become important in
region II, however, since the solution is a flat space orbifold it
seems reasonable to expect that such corrections vanish.

In the above analysis no assumption was made regarding the size of the
compactification radius $R$ comparing with the string
length. For some space--like surface at time $t$ in region $I$, 
provided $\D$ is sufficiently small, the typical energy scale $E=\D/R$
for phenomena on the cosmological solution is always much smaller than
the Ka\l u\.{z}a--Klein and string mass gaps. Also, the
nine--dimensional string coupling is small.  In
region $II$ the Ka\l u\.{z}a--Klein states become very massive but the
associated electric field becomes very large pointing to the
Schwinger process, the winding states become very light and the
nine--dimensional string coupling blows up. Clearly in region $II$ the
relevant description is in term of the ten--dimensional string
orbifold, to which we now turn.

\subsection{String partition function}

In this section we start to analyze the motion of a (bosonic) string in the
quotient space described in the previous sections. We leave the
generalization to the superstring to future work. Readers that are not
familiar with string theory may wish to skip this subsection and
to move to the classical gravity analysis in section 4 (references
\cite{GSW,Polch} provide the necessary background for the
techniques here presented). Similar computations have been carried out
in the case of open strings in
electric fields \cite{BachPorr:92}, D--branes in relative motion
\cite{Bachas:95} and closed strings in magnetic backgrounds
\cite{RussoTsey:94,Kiritsis:95}. We will limit ourselfs to the
computation of the one--loop partition function, leaving to future
work a more detailed analysis of the results. We will use units such
that $\alpha ^{\prime }=2$.

We use, in the following, lightcone coordinates 
\eqn
X^{\pm }=\frac{1}{\sqrt{2}}\left( X\pm T\right)\ ,
\eeqn
so that the basic identifications introduced in section $2$ are given by 
\eqn
X^{\pm }\sim e^{\pm 2\pi \Delta }X^{\pm }\ ,
\ \ \ \ \ \ \ \ \ \ \ \ \ \ \ 
Y\sim Y+2\pi R\ .  
\label{eq100}
\eeqn
Let us then focus on the winding sector with winding number $w$. First of all, 
it is clear that the
mode expansion of the field $Y\left( z,\overline{z}\right)$ 
is the usual one of a compact boson. The only difference with
the standard $S^{1}$ compactification is given by a modified constraint on the
total momentum $P$, which must be compatible with the identification (\ref
{eq100}) and must therefore satisfy 
\eqn
\exp{\left[2\pi i\left(RP+\Delta J\right)\right]}=1\ ,
\ \ \ \ \ \ \ \ \ \ \ \ \ \ 
P=\frac{1}{R}\left( n-\Delta J\right)\ ,
\eeqn
where $n$ is an integer and $J$ is the boost operator.
The left and right momenta for $Y$ are then given by 
\eqn
p_{L,R}=P\pm \frac{wR}{2}\ .
\eeqn
The mode expansions of the fields $X^{\pm }\left( z,\overline{z}\right) $
are, on the other hand, modified and are given explicitly by 
\eqn
X^{\pm }\left( z,\overline{z}\right) =i\sum_{n}
\left(\frac{1}{n\pm i\nu }
\frac{a_{n}^{\pm }}{z^{n\pm i\nu }}+\frac{1}{n\mp i\nu }
\frac{\widetilde{a}_{n}^{\pm }}{\overline{z}^{n\mp i\nu }}\right)\ ,
\eeqn
where $\nu =w\Delta $ and where the oscillators satisfy the commutation
relations 
\eqn
\left[ a_{m}^{\pm },a_{n}^{\mp }\right] =
\left( m\pm i\nu \right) \delta_{m+n}\ ,
\ \ \ \ \ \ \ \ \ \ \ \ \ \ \ 
\left[ \widetilde{a}_{m}^{\pm},\widetilde{a}_{n}^{\mp }\right] 
=\left( m\mp i\nu \right) \delta _{m+n}\ ,
\eeqn
and the hermitianity conditions 
$\left( a_{m}^{\pm }\right)^{\dagger}=a_{-m}^{\pm }$, 
$\left(\widetilde{a}_{m}^{\pm }\right)^{\dagger }
=\widetilde{a}_{-m}^{\pm }$. The contribution to the
Virasoro generators from the fields $X^{\pm }$
has been computed previously in \cite{Bachas:95,RussoTsey:94} and is given by 
($\cdots$ denotes contributions from other fields)
\eqn
\arr{l}
\displaystyle{L_{0}=\cdots+
\frac{1}{2}i\nu \left( 1-i\nu \right)+
\sum_{n\geq1}a_{-n}^{+}a_{n}^{-}+
\sum_{n\geq 0}a_{-n}^{-}a_{n}^{+}}\ ,
\spa{0.5}\\
\displaystyle{\widetilde{L}_{0}=\cdots+
\frac{1}{2}i\nu \left( 1-i\nu \right)+
\sum_{n\geq 0}\widetilde{a}_{-n}^{+}\widetilde{a}_{n}^{-}+
\sum_{n\geq 1}\widetilde{a}_{-n}^{-}\widetilde{a}_{n}^{+}}\ ,
\earr
\eeqn
where implicitly we have chosen to consider $a_{0}^{-}$ and 
$\widetilde{a}_{0}^{+}$ as creation operators\footnote{The correct
treatment of the zero--modes is quite subtle. 
In fact they form two copies of the Heisenberg algebra $[a_0^+,a_0^-]=
[\widetilde{a}_0^-,\widetilde{a}_0^+]=i\nu$. We choose to quantize 
the zero--modes as creation and annihilation operators in order to
preserve the conformal invariance of the partition function.}.
Finally, the boost operator $J=J_{L}+J_{R}$ defined by 
$i\left[J,X^{\pm}\right]=\pm X^{\pm}$, is given explicitly by 
\eqn
J_{L}=-i\sum_{n\geq 1}N_{n}^{+}+i\sum_{n\geq 0}N_{n}^{-}\ ,
\ \ \ \ \ \ \ \ \ \ \ \ \ \ 
J_{R}=-i\sum_{n\geq 0}\widetilde{N}_{n}^{+}
+i\sum_{n\geq1}\widetilde{N}_{n}^{-}\ ,
\eeqn
where $N_{n}^{\pm}=\left(n\mp i\nu\right)^{-1}a_{-n}^{\pm}a_{n}^{\mp}$
and $\widetilde{N}_{n}^{\pm}=
\left(n\pm i\nu\right)^{-1}\widetilde{a}_{-n}^{\pm}\widetilde{a}_{n}^{\mp}$ 
are the usual occupation numbers. It is then clear that one can
rewrite the \textit{total} Virasoro generators for the three bosons
$T$, $X$, and $Y$ in terms of the usual 
\textit{integral }level numbers 
$\bL,\widetilde{\bL}$ 
and the boost operator as 
\eqn
\arr{l}
\displaystyle{L_{0} =
\frac{1}{2}\,i\nu \left( 1-i\nu \right)+
\nu J_{L}+\frac{1}{2}\,p_{L}^{2}+\bL}\ , 
\spa{0.5}\\
\displaystyle{\widetilde{L}_{0} =
\frac{1}{2}\,i\nu \left( 1-i\nu \right) -\nu J_{R}+
\frac{1}{2}\,p_{R}^{2}+\widetilde{\bL}}\ .
\earr
\eeqn

We are now ready to compute the partition function
$\;Z_{3}\,\ $for the three bosons $T$, $X$ and $Y$.
We have 
\eqn
\arr{rcl}
\displaystyle{Z_{3}=
\left(q\overline{q}\right)^{-\frac{1}{8}}}
&\displaystyle{\sum_{w,n}}&
\displaystyle{{\mathrm{Tr}}\,q^{\bL}\,\overline{q}^{\widetilde{\bL}}
\left( \frac{q}{\overline{q}}\right) ^{\frac{1}{2}nw}
\left( q\overline{q}\right) ^{\frac{1}{2}
\left[\left( \frac{wR}{2}\right) ^{2}+
\left( \frac{n-\Delta J}{R}\right) ^{2} \right]}} 
\spa{0.6}\\&&
\displaystyle{\times
\left(q\overline{q}\right)^{\frac{1}{2}\nu\left(J_{L}-J_{R}\right)}
\left(q\overline{q}\right)^{\frac{1}{2}i\nu\left(1-i\nu\right)}}\ ,
\earr
\eeqn
where, as costumery, $q=e^{2\pi i\tau }$ and 
$\tau =\tau _{1}+i\tau_{2}$. Performing the usual Poisson 
resummation on $n$ brings the above expression to the simpler form 
\eqn
\arr{rcl}
\displaystyle{Z_{3} = 
\left(q\overline{q}\right)^{-\frac{1}{8}}\frac{R}{\sqrt{2\tau_{2}}}}
&\displaystyle{\sum_{w,w^{\prime }}}&
\displaystyle{\exp\left[-\frac{\pi R^{2}}{2\tau _{2}}T\overline{T}
-2\pi \tau _{2}\Delta^{2}w^{2}\right]} 
\spa{0.6}\\&&
\displaystyle{\times\, q^{\frac{1}{2}i\nu}\,{\mathrm{Tr}}_{L}
\left(e^{2\pi i T\Delta J_{L}}q^{\bL}\right) 
\overline{q}^{\frac{1}{2}i\nu }\,{\mathrm{Tr}}_{R}
\left( e^{2\pi i \overline{T}\Delta J_{R}}
\overline{q}^{\widetilde{\bL}}\right)}\ ,
\earr
\eeqn
where
\eqn
T=w\tau -w^{\prime }\ .
\eeqn
In the above sum the term with $w=w^\prime=0$, which is by
itself modular invariant, gives the usual 
partition function of the uncompactified theory. We will therefore focus,
in the following, only on the other terms in the sum (we will denote the
restricted sum by $\sum^\prime$). With this in mind,  
the computation of the holomorphic trace $\mathrm{Tr}_{L}$ is
simple. Defining 
$c=e^{2\pi i\left(i\Delta T\right)}=q^{i\nu }e^{2\pi w^{\prime}\Delta }$, 
one has 
\eqn
\arr{rcl}
\displaystyle{{\mathrm{Tr}}_{L}
\left(e^{2\pi i T\Delta J_{L}}q^{\bL}\right)}
&=&
\displaystyle{
\frac{1}{1-c}\prod_{n\geq 1}\frac{1}{\left( 1-q^{n}\right) 
\left( 1-q^{n}c\right) \left( 1-q^{n}c^{-1}\right)}}
\spa{0.6}\\&=&
\displaystyle{iq^{\frac{1}{8}}c^{-\frac{1}{2}}\frac{1}{\theta _{1}
\left( i\Delta T|\tau\right) }}\ .
\earr
\eeqn
Similarly, the antiholomorphic trace is given by 
$\mathrm{Tr}_{R}=\overline{c}\,\overline{\mathrm{Tr}_{L}}$. 
Therefore, the partition function $Z_{3}$ is given by the
final expression (reinserting $\alpha^{\prime}$) 
\eqn
Z_{3}=\frac{R}{\sqrt{\alpha^\prime \tau_2}}
{\sum_{w,w^{\prime }}}^\prime
\ e^{-\frac{\pi R^{2}}{\alpha ^{\prime}}
\frac{T\overline{T}}{\tau _{2}}-2\pi\tau_{2}\Delta^{2}w^{2}}
\left| \theta_{1}
\left(i\Delta T|\tau\right)\right|^{-2}\ .
\label{eq200}
\eeqn
Using the fact that $\partial_{\zeta}\theta_{1}\left(\zeta |\tau\right)|_{\zeta =0}
=2\pi \eta ^{3}\left(\tau\right)$, we see that, as
$\Delta \to 0$, the partition function diverges. This is due to
the fact that, in the limit, the rotational symmetry in the $X$--$T$
plane is restored and the partition function is then proportional 
to the space--time volume.
Moreover, the modular 
properties of $\theta _{1}$ insure modular invariance of the full expression.

In order to compute the total partition function, we can use the usual 
trick \cite{Polchinski:86} of extending the integration region from
the fundamental domain $|\tau|>1$, $|\tau_1|<\frac{1}{2}$ to the
strip 
$\Gamma=\{\tau\in\bC|\tau_{2}>0,
\left|\tau_{1}\right|<\frac{1}{2}\}$, 
while at the same time restricting the sum in (\ref{eq200}) to 
$w=0,w^{\prime }\geq 1$. Then we have 
\eqn
Z=\int_{\Gamma }
\frac{d\tau d\overline{\tau }}
{\tau _{2}^{2}}\,Z_{g}\,Z_{b}^{23}\,\widetilde{Z}_{3}\ ,  
\label{eq300}
\eeqn
where $Z_{g}$ and $Z_{b}$ are the usual partition functions for the 
$b$--$c$ ghosts and for a non--compact boson, and where  
\eqn
\widetilde{Z}_{3}=
\frac{R}{\sqrt{\alpha^\prime \tau_2}}
\sum_{w^{\prime }\geq 1}e^{-\frac{\pi R^{2}}{\alpha^{\prime}}
\frac{w^{\prime 2}}{\tau _{2}}}
\left|\theta_{1}
\left( i\Delta w^{\prime}|\tau\right)\right|^{-2}\ .
\eeqn

Let us note that the above expression has strong similarities with the
expression for the Euclidean BTZ black hole partition function found 
in \cite{Maldacena:00}, as expected from the analysis in subsection
2.2. The expression for $\widetilde{Z}_{3}$ in fact
exhibits a similar structure of poles in the region $\Gamma $, which are in
correspondence with the zeros of $\theta _{1}\left( i\Delta w^{\prime }|\tau
\right) $. The function $\theta _{1}\left( \zeta |\tau \right) \,$, as a
function of $\zeta $, has in fact a simple zero at the points 
$\zeta =a\tau-b\,$($a,b\in \bZ$). More precisely 
\eqn
\theta_{1}\left(\zeta|\tau\right)\sim\left(-\right)^{a+b}2\pi\, 
\eta^{3}\left( \tau \right) e^{-\pi ia\left( a+1\right) \tau }
\left(\zeta-a\tau+b\right)\ .
\eeqn
Then the poles of $\widetilde{Z}_{3}$ are determined by the equation 
$i\Delta w^{\prime}=a\tau -b$ and are located at
\eqn
\tau =\frac{1}{a}\left( b+i\Delta w^{\prime }\right)\ .
\eeqn
In order for the poles to be in $\Gamma $, we must have $a\geq 1$ and 
$\left| b\right| \leq 2a$. For fixed $w^{\prime }$, the structure of the
poles is the same as those found in \cite{Maldacena:00}.

Finally let us comment on the origin of the poles. They arise from the 
zero--modes of the winding sector. In fact, stretched winding strings
have a length which is $2\pi R$ when they sit at the origin $X=T=0$,
but which decreases as they approach the causal singularity, where
they become almost massless. One can easily analyse the
classical dynamics of these winding strings. In particular, in region
$I$ these strings move from the collapsing to the expanding regions,
either passing through the origin where they have minimal length or
passing through the intermediate region. At large
cosmological times these strings become very long and stretched. Such
states, could be the origin of the pole structure of $Z_3$. In
particular, the physical meaning of $Z_3$ could be related to
nucleation processes of such long strings.

We leave to future work a more detailed analysis of the integral
(\ref{eq300}) and of the precise physical meaning of the poles in 
$\widetilde{Z}_{3}$, and the study of the possible relation of the
winding string modes with the long strings  studied in
\cite{Maldacena:98,Seiberg:99,Maldacena:00b,Maldacena:00}.

\subsection{M--Theory and 9--11 flip}

In this subsection we comment on the M--theory compactification. Consider
the construction of section two and add eight spectator flat
directions. Then reducing to the IIA theory one obtains the following
background fields 
\eqn
\arr{c} 
\displaystyle{ ds_{10}^{\ \ 2}=\La^{1/2}\left[-dt^2+ds^2(\bE^8)\right]
+\frac{(Et)^2}{\La^{1/2}}\,dx^2\ ,}
\spa{0.4}\\
\displaystyle{ e^{4\phi/3}= \La(t)\ ,\ \ \ \ \ 
A=\frac{(Et)^2}{\La}\,dx\ .}  
\earr
\label{Mthsol}
\eeqn 
By compactifying one of the spectator directions on a circle, this 
solution is related to the previous nine--dimensional solution by a
$9-11$ flip. Also, one can analyse the validity of the gravity
approximation, finding that this description is appropriate for a
large expansion period in region $I$ provided that the boost parameter
$\D$ is sufficiently small, and that the compactification radius $R$
satisfies $R\ll\sqrt{\a'}$ (and therefore $g\ll 1$). In region $II$, the
string coupling becomes small since the proper length of the
compactification circle is decreasing. Near the causal singularity
the string coupling becomes zero and the compactification radius becomes
light--like. Hence one expects that the correct description of the
apparent ten--dimensional singularity is given by Matrix theory. We
think this issue deserves further investigation because it would give a  
resolution of the singularity in the above cosmological solution.

\section{Cosmological Solutions in Arbitrary Dimensions}
\news

So far we have considered geometries that arise from
Ka\l u\.{z}a--Klein compactifications. Naturally, one expects that such
geometries can be generalized to arbitrary dimensions and arbitrary
degree of the form gauge field. These new space--times should
share many properties with the case where space--time is
a quotient of flat space. The situation is analogous to the
BTZ black hole, which is simply a quotient of three-dimensional
Anti--De Sitter space but retains the standard properties of higher
dimensional black holes.

A short cut to construct such generalization is to realize the
similarity between the dilatonic Melvin solution
\cite{Gibbons:86} and the cosmological solution of section 2. 
In the case of the Ka\l u\.{z}a--Klein Melvin solution, the geometry 
is simply flat--space with an
identification along the orbits of the isometry subgroup generated by
rotations on a plane together with translations
\cite{Dowker:93}. Hence, replacing the boost by a rotation in our
construction one recovers the Ka\l u\.{z}a--Klein Melvin solution. 
For example, starting with the cosmological solution (\ref{Mthsol})
and making the analytic continuation $t\rightarrow ir$, 
$E\rightarrow iE$, $\bE^8\rightarrow \bM^8$, one obtains the 
flux 7--brane solution of the type IIA theory
\cite{RussoTsey,CostaGutp:00}. This suggests that the
generalization of the two--dimensional cosmological solution to higher
dimensions can be found by an analytic continuation of the flux brane
geometries \cite{Saffin:01,GutpStro:01,us}. Earlier work on
cosmological string backgrounds can be found in  
\cite{Myers:87,Antoniadis:88,TseytlinVafa:91,Tseytlin:91}.

\subsection{The action and basic ansatz}

To keep our discussion general we shall consider a D--dimensional
space--time with a $\tilde{d}$-form field strength $F=dA$
and a scalar field $\phi$. The corresponding gravitational action is 
\eqn 
S=\frac{1}{2\kappa_D^{\ 2}}\int d^Dx\sqrt{-g}
\left[R-\frac{1}{2}(\partial\phi)^2 
-\frac{1}{2\tilde{d}!}e^{\a\phi}F^2\right]\ , 
\label{action} 
\eeqn 
where $\kappa_D$ is the gravitational coupling. Of course this action
can be regarded as a consistent truncation of either String or
M--theory low energy actions, where $F$ represents any of the
field strengths or electromagnetic dual in these theories. We shall
reduce the theory to $d+1=D-\tilde{d}$ dimensions according to the ansatz
\eqn
\arr{c}
\displaystyle{ds_D^{\ 2}=e^{-\frac{2\tilde{d}}{d-1}\,\la}ds^{2}
+e^{2\,\la}ds^2\left(\bE^{\tilde{d}}\right)}\ ,
\spa{0.3}\\
\displaystyle{F=E\,\e\left(\bE^{\tilde{d}}\right)}\ .
\label{KKred}
\earr
\eeqn
Then the action (\ref{action}) becomes effectively
\eqn
S=\frac{1}{2\kappa^2}\int d^{d+1}x\sqrt{-g}
\left[R-\frac{1}{2}(\partial\phi)^2
-\frac{\tilde{d}(\tilde{d}+d-1)}{d-1}(\partial\la)^2
-V(\phi,\la)\right]\ ,
\label{KKaction} 
\eeqn 
where $\kappa$ is the $(d+1)$--dimensional gravitational coupling 
and we conveniently use the Einstein metric. The
potential $V(\phi,\la)$ has the form
\eqn
V(\phi,\la)=\frac{E^2}{2}\,
\exp{\left(\a\,\phi-2\frac{d\tilde{d}}{d-1}\,\la\right)}\ .
\label{potential}
\eeqn

Before deriving the equations of motion for the action it is
convenient to define the scalar fields $\rho$ and 
$\psi$ by the relations
\eqn
\arr{c}
\displaystyle{\rho=\a(\tilde{d}+d-1)\,\la+d\,\phi}\ ,
\spa{0.2}\\
\displaystyle{\psi=\frac{2d\tilde{d}}{d-1}\,\la-\a\,\phi}\ .
\earr
\label{newscalars}
\eeqn
With this field redefinition, $\rho=\rho_0$ constant solves the
equation of motion for this scalar, which decouples from the remaining
field equations. Then the gravitational action coupled to the scalar $\psi$
becomes 
\eqn
S=\frac{1}{2\kappa^2}\int d^{d+1}x\sqrt{-g}
\left[R-\frac{\b}{2}(\partial\psi)^2
-V(\psi)\right]\ ,
\label{KKaction2} 
\eeqn 
where the potential $V(\psi)$ has the form
\eqn
V(\psi)=\frac{E^2}{2}\,\exp{(-\psi)}\ ,
\label{potential2}
\eeqn
and $\b$ is the numerical factor
\eqn
\b^{-1}=\a^2+\frac{2d^2\tilde{d}}{(\tilde{d}+d-1)(d-1)}\ .
\label{beta}
\eeqn

\subsection{Region $I$}

To find the metric and scalar field in the region that is analogous to
the  region $I$ of the Ka\l u\.{z}a--Klein case, consider the
Robertson--Walker space--time metric for an open universe  
\eqn
\arr{rcl}
ds^{2}&=&
\displaystyle{-dt^2+a^2(t)\,ds^2(H_d)}
\spa{0.4}\\&=&
\displaystyle{-dt^2+a^2(t)
\left(d\chi^2+\sinh^2\chi\,d\O^{\ 2}_{d-1}\right)}\ ,
\earr
\label{RW}
\eeqn
together with the time-dependent scalar field $\psi(t)$.
The equation of motion for the scalar is
\eqn
\b\left(\ddot{\psi}+d\,\frac{\dot{a}}{a}\,\dot{\psi}\right)=
-\frac{\p V}{\p\psi}=V(\psi)\ ,
\label{scalarseqn}
\eeqn
while Einstein equations take the usual form
\eqn
\arr{rcl}
\displaystyle{\left(\frac{\dot{a}}{a}\right)^2-\,\frac{1}{a^2}}
&=&
\displaystyle{\frac{1}{d(d-1)}
\left[\frac{\b}{2}\,\dot{\psi}^2+V(\psi)\right]}\ ,
\spa{0.7}\\
\displaystyle{\frac{d\,}{dt}\left(\frac{\dot{a}}{a}\right)+\frac{1}{a^2}}
&=&
\displaystyle{\frac{\b}{2(1-d)}\,\dot{\psi}^2}\ ,
\earr
\label{Friedmann}
\eeqn
where dots represent derivatives with respect to the coordinate $t$.

In analogy with the two--dimensional case we look for a geometry with
similar behavior around $t=0$. In particular, we require the geometry
to have the Milne form along the light rays. This is only possible
provided the homogeneous space is the hyperboloid $H_d$, justifying
our choice for an open universe cosmology.
Then, one can find a solution to the above system of differential
equations as a power expansion in the dimensionless quantity $(Et)$.
A straightforward calculation gives for the first terms in this
expansion
\eqn
\arr{c}
\displaystyle{a(t)=t\,
\left(1+\frac{e^{-\psi_0}}{3d(d-1)}\,(Et)^2+\cdots\right)}\ ,
\spa{0.5}\\
\displaystyle{\psi(t)=\psi_0+\frac{e^{-\psi_0}}{4\b}\,(Et)^2+\cdots}\ .
\earr
\label{nearsol}
\eeqn

Next we want to find the asymptotics
for this geometry. As it is the case with cosmological solutions in an
open universe, at later times we have curvature domination. Hence,
the correct ansatz for the asymptotic solution is to set $a=a_0\,t$, and
then solve for the scalar field $\psi(t)$ and use the Freedman equation
to fix the constant $a_0$. This results in the asymptotic behavior
\eqn
a(t)=a_0\,t\ ,\ \ \ \ \ \ \ 
e^{\psi(t)}=\frac{(Et)^2}{4(d-1)\b}\ ,
\label{asyptsol}
\eeqn
with the constant $a_0$ given by
\eqn
a_0^{\ 2}=\left(1-\frac{2\b}{d-1}\right)^{-1}\ .
\label{a_0}
\eeqn

The picture we have in region $I$ is exactly the same as for the 
Ka\l u\.{z}a--Klein case. An observer that is comoving with the
expansion could think that at an earlier time the matter density
blows up and there is a cosmological space--like
singularity. This is the usual understanding in cosmology. However, in
the picture we are proposing here, there is a past cosmological
horizon where the geometry is perfectly smooth. In fact, we have fixed
the boundary conditions on this horizon to evolve into the
future. Of course, there is also a collapsing region $I$, where the
comoving observer sees a future horizon where the boundary conditions
are imposed. We shall give a more detailed discussion of this proposal
in the conclusion. 

\subsection{Region $II$}

The boundary conditions imposed above on the surface $t=0$,
$\chi=+\infty$ are that the geometry looks like the Milne
universe, which in fact is just flat space. The
non--trivial assumption we are making is that the scalar field
and the conformal factor have the same behavior through out the whole surface $t=0$, 
$\chi=+\infty$. This is essential to have a homogeneous
cosmology. 

We can pass from the expanding (or collapsing) region $I$ to a
region $II$
by a suitable change of coordinates. Again there is a close analogy
with the two--dimensional case. First we change to a coordinate system
well behaved around the coordinate singularity at $t=0$
\eqn
\tilde{T}=t\cosh\chi\ ,\ \ \ \ \ \ 
\tilde{X}=t\sinh\chi\ .
\eeqn
In these coordinates the metric and scalar fields around the light rays
$|\tilde{T}|=|\tilde{X}|$ are well behaved and there is no coordinate
singularity, which allows us to continue the solution to region $II$ 
where $|\tilde{T}|<|\tilde{X}|$. To make the symmetries of space
manifest it is convenient to define new $(t,\chi)$ coordinates in
region $II$ by
\eqn
\tilde{T}=t\sinh\chi\ ,\ \ \ \ \ \ 
\tilde{X}=t\cosh\chi\ .
\eeqn
Then the metric ansatz takes the form 
\eqn
\arr{rcl}
ds^{2}&=&
\displaystyle{+dt^2+a^2(t)\,ds^2(dS_d)}
\spa{0.4}\\&=&
dt^2+a^2(t)\left(-d\chi^2+\cosh^2\chi\,d\O^{\ 2}_{d-1}\right)\ .
\earr
\label{RW2}
\eeqn
The $SO(1,d)$ symmetry of the original ansatz in region $I$, realized on the
constant time $d$--dimensional hyperboloids, is now the symmetry of the   
constant $t$ space--time slices. In fact, these slices are simply the
$d$--dimensional De Sitter space. Notice that, in region $II$, the
coordinate $t$ becomes space--like and the coordinate $\chi$ time--like.
Along the horizon the $SO(1,d)$ symmetry acts as translations
justifying the boundary conditions we have imposed. 

The form of the solution near the $t=0$ surface can be obtained simply
by the analytic continuation of the solution in region $I$
\eqn
a_{II}(t)=-ia_I(it)\ , 
\ \ \ \ \ 
\psi_{II}(t)=\psi_I(it)\ .
\eeqn
This gives the following expansion
\eqn
\arr{c}
\displaystyle{a(t)=t\,
\left(1-\frac{e^{-\psi_0}}{3d(d-1)}\,(Et)^2+\cdots\right)}\ ,
\spa{0.5}\\
\displaystyle{\psi(t)=\psi_0-\frac{e^{-\psi_0}}{4\b}\,(Et)^2+\cdots}\ .
\earr
\label{nearsolII}
\eeqn

\begin{figure}[ht]
\begin{picture}(0,0)(0,0)
\put(80,247){${\cal I}^+$}
\put(230,162){${\cal I}^+$}
\put(80,50){${\cal I}^-$}
\end{picture}
\centering\psfig{figure=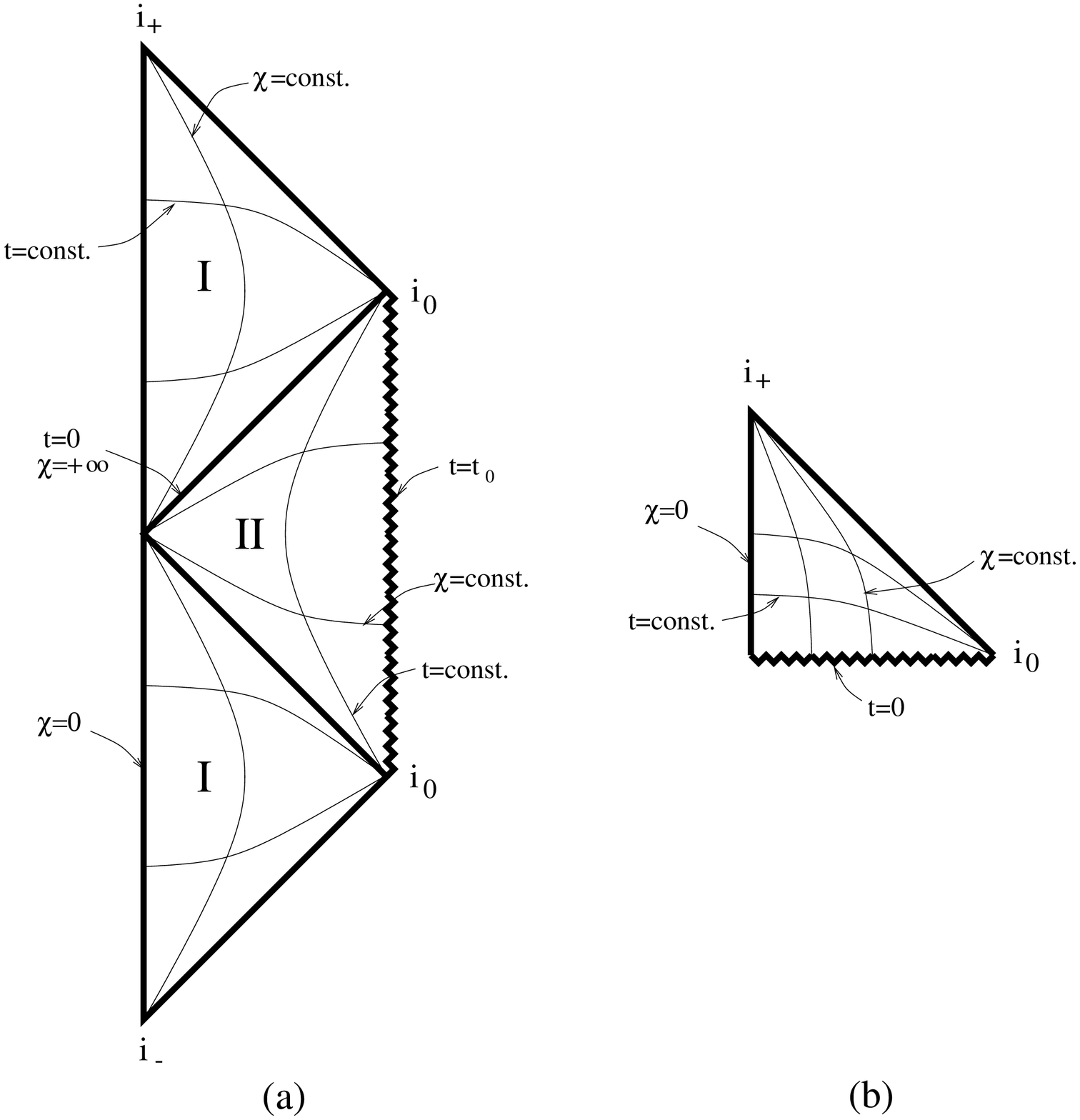,width=10cm} 
\caption{\small{Carter--Penrose diagrams for open Universe
cosmologies. In both diagrams each point represents a $(d-1)$ sphere
and $\chi=0$ is a coordinate singularity. The standard diagram in (b)
is presented for comparison with our proposal.}} 
\label{fig3}
\end{figure}

To find the asymptotics of the solution in region $II$, notice that if
$a(t)=\cos{t}$ the metric describes $(d+1)$--dimensional De Sitter space
and the surface $t=\pi/2$ is a coordinate singularity. In fact, if
this was the case for this solution we would have a cosmological
solution without any singularity. Starting from region $II$ at $t=\pi/2$
we would have a De Sitter phase, which would then evolve to the
Robertson--Walker cosmology in region $I$. Unfortunately this is not the
case for the potential $V(\psi)$ that results from the Ka\l u\.{z}a--Klein
compactification. We have done some numerics and verified that the
solution develops a singularity around  some $t=t_0$, exactly as in
the two--dimensional case. The correct asymptotics can be found by
doing the ansatz
\eqn
\arr{c}
\displaystyle{a(t)=a_0\,(t_0-t)^{\gamma}+\cdots}\ ,
\spa{0.2}\\
\displaystyle{\psi(t)=\eta\log{\left(\theta E(t_0-t)\right)}+\cdots}\ .
\earr
\label{asymptII}
\eeqn
Then the equations of motion give the following values for the
dimensionless constants $\eta$, $\gamma$ and $\theta$ 
\eqn
\eta=2\ ,\ \ \ \ \ \ 
\gamma=\frac{2\beta}{d-1}\ ,\ \ \ \ \ \ 
\theta^{-2}=4\b\left(1-d\,\gamma\right)\ ,
\eeqn
while the value of $a_0$ is a constant of integration, that is fixed
by the value of the scalar field at the light rays. 

We have verified that the analytic behavior for the different asymptotics in both
regions $I$ and $II$ exactly matches the numerical
analysis. In figure 3a the Carter--Penrose diagram for the
cosmological solutions is shown. For comparison we included in figure
3b the usual Robertson--Walker open Universe diagram. The pre
Big--Bang  string cosmology scenario \cite{Veneziano:00}, considers
the latter diagram and glues a collapsing phase to an expanding phase
along the space--like singularity. For the new space-time global
structure here presented $t=0$ is a cosmological horizon. For small
$t$ the universe undergoes an acceleration and, when it becomes
curvature dominated, it expands linearly with time.

\subsection{String and M--theory cosmologies}

As mentioned before the above solutions are naturally embedded in
String and M--theory. So we would like to understand where the gravity
approximation is valid and where space--time and world--sheet string
effects become important. The behavior of the solution is
essentially determined by the constant $\a$ in the coupling between
the dilaton field and the gauge field. For the RR gauge fields of the
type II theories we have $\a=(5-\tilde{d})/2$, for the NS--NS 2--form gauge 
field $a=-1$ and for its dual 6--form gauge field $a=1$. If 
$\a>0$ ($\a<0$) the string coupling will become very large (small)
near the singularity and it will become very small (large) in the
asymptotics of regions $I$. In all cases the volume of the compact space
becomes very small near the curvature singularity, which is telling us that the
string winding states become very light. Asymptotically space--time
decompactifies. 
 
Let us analyse in more detailed the cases with $\a=0$. Consider first
the IIB theory with the self--dual 5--form field strength. As for the
D3--brane the string coupling can be made small throughout the whole
solution. There will be curvature corrections near the singularity
where string winding states become very light and the gravity
approximation breaks down. Two other cases with $\a=0$ correspond
to the M--theory 4--form field strength or its dual 7--form. The latter
case is particularly interesting. For $\tilde{d}=7$,
$\a=0$ and $D=11$, we have a four--dimensional cosmology for an open
universe, and the compact manifold is seven-dimensional. Through out
this paper we considered flat manifolds in the internal directions, 
but more generally one can consider any Ricci flat manifold. In this
case, one could consider a $G_2$ manifold for which there has been a
great deal of interest for flat space compactifications, which yield
${\cal N}=1$ SUSY in $D=4$ (see \cite{Atiyah:01} and references there
in). Here, supersymmetry is broken by the form flux \cite{StroPolc:95}. 

As this work was being completed an interesting paper
\cite{GutpStro:02} appeared,
also considering new time dependent string theory backgrounds. These
backgrounds are different from the ones introduced here. They
analyzed the case where the boundary conditions for the scalar
field are set on a spacelike hypersurface, while we have considered
the case of a null hypersurface. We have interpreted this null
hypersurface as the cosmological horizon in an open Universe
cosmology. This fact allowed us to continue the geometry
in both directions from the horizon. A natural question is whether
one can also continue the other new geometries presented in \cite{GutpStro:02}
to a region similar to region $II$ here described. For the 
$D=4$ Einstein--Maxwell solution of \cite{GutpStro:02}, it is
indeed possible to extend the geometry to a region $II$ where there
will be a time--like singularity. The Penrose diagram is similar to
the diagram of figure 2 with each point representing a
two--dimensional hyperbolic plane.

\section{Conclusion: Towards a Solution of the Horizon
\\ and Singularity Problems?}
\news

In this paper we have proposed a new cosmological scenario that tries
to evade the cosmological singularity problem. The essential point
was to consider collapsing and expanding phases with future and past
cosmological horizons, and to fix the boundary conditions along these
null hypersurfaces. From these horizons,
we constructed the space--time geometry for specific examples
that arise quite naturally in String theory. In the cosmological solution
that we have studied, and more generally in cosmological solutions that
have identical space-time causal structure, the usual horizon problem
of standard cosmology does not arise. In fact, consider two points
of the hyperboloid at constant cosmological time $t$ in the expanding
phase, that have an arbitrarily large space--like separation. Then from
the Carter--Penrose diagram, because there is a past cosmological horizon,
the past light cones of these points always intersect.

Another aspect of this proposal is that it rises the hope of
resolving the cosmological singularity problem. Firstly, in the
two--dimensional toy model here presented we were able, by embedding
the geometry in string theory, to interpret the geometry as arising
from a string theory orbifold from which one can at least do some
calculations. In particular, we have started the analysis of the 
one--loop partition function which is modular invariant.
The analogy with strings in thermal $AdS_3$ \cite{Maldacena:00}, is
definitely worth pursuing. This toy model was inspired by the recent
proposal in \cite{Khoury:01}, and provides a regulator for their
proposal. Indeed, with the identifications by a boost and a
translation,  at small translation parameter $R$ space--time is still
smooth (in particular Hausdorff) and there is a horizon.
So this looks a like good regulator for the singular space--time one
obtains when setting $R=0$ from the beginning. Secondly, in the case
of the higher dimensional solutions, there are cases where the string
coupling can be kept small through out the whole space--time. There is a
time--like singularity in the region $II$ of space--time, where
the compact space is shrinking and where winding string states become
important. It would be very interesting to resolve this singularities
within string theory. A reasonable conjecture is that the new light
degrees of freedom could acquire some VEV resolving the singularity. 

There is possibly another way of avoiding the cosmological
singularity. This could happen if the potential for the scalar field
had a different shape. Then one would start from a pure De Sitter
phase in region $II$, as the fields evolve into the horizon one
would be able to smoothly pass to the expanding region $I$. We have
tried some numerics with different types of potentials but so far have
not succeed. Essentially, there is a fine--tuning problem. As one starts from
the De Sitter phase, it seems very difficult to evolve the differential
equations such that we reach the horizon with the exact behavior for
the scale factor ($\dot{a}=1$), and scalar field ($\dot{\psi}=0$), such
that the transition can be achieved. On the other hand, there may
exist solitons that interpolate between different vacua in regions
$I$ and $II$ (like domain--walls). We think this issue deserves
further attention because, if successful, it would give a completely smooth
cosmological solution. Let us note that this is not in contradiction
with the singularity theorems because in the De Sitter phase the
strong energy condition is not satisfied.

Finally, independently of string theory, one should revisit many issue
in cosmology by considering different types of potentials, as those
used in inflationary and particle cosmology,  and by imposing
the boundary conditions at the cosmological horizon. From these
boundary conditions the universe can evolve into the usual
cosmological epochs. Another important issue that we plan to analyse
is the question of thermal radiation seen by the cosmological
observer. Consider, for example, a scalar field and fix its boundary
conditions to have only positive frequency modes on the
cosmological horizon. Then one can study the spectrum at later
times. Note that the same reasoning cannot be applied in the
presence of a space--like Big--Bang singularity.

\section*{Acknowledgments}
We are grateful to Costas Bachas, Carlos Herdeiro, Costas Kounnas and
Rodolfo Russo for helpful discussions. The work of M.S.C. was
supported by a Marie Curie Fellowship under the European Commission's
Improving Human Potential programme. L.C. would like to thank the
Newton Institute in Cambridge where this work was completed.
This work was partially supported by CERN under contract 
CERN/FIS/43737/2001.

\end{document}